\documentclass[aps,prd,preprint,showpacs,showkeys,superscriptaddress]{revtex4}

\usepackage{latexsym}
\usepackage{amsmath,amssymb}
\usepackage[dvips]{graphicx}
\usepackage[dvips]{hyperref} 

\begin{document}


\title{Smooth cosmological phase transition in the  Ho\v{r}ava-Lifshitz gravity}

\author{Edwin J. Son}
\email[]{eddy@sogang.ac.kr}
\affiliation{Center for Quantum Spacetime, Sogang University, Seoul 121-742, Korea}

\author{Wontae Kim}
\email[]{wtkim@sogang.ac.kr}
\affiliation{Center for Quantum Spacetime, Sogang University, Seoul 121-742, Korea}
\affiliation{Department of Physics, Sogang University, Seoul 121-742, Korea}
\affiliation{School of Physics, Korea Institute for Advanced Study, Seoul 130-722, Korea}

\date{\today}

\begin{abstract}
We show that the cosmological phase transition from the first
accelerated expansion in the early universe to the second accelerated
expansion over the intermediate decelerated
expansion is possible in the HL gravity without the ``detailed balance'' condition if the \emph{dark scalar}
energy density is assumed to be negative. 
Moreover, we obtain various evolutions depending on the scale
factor and the expansion rate. Finally, we discuss the
existence of the minimum scale in connection with the singularity free 
condition. 
\end{abstract}

\pacs{04.70.Dy, 04.60.Kz}

\keywords{gravitational Chern-Simons, entropy}

\maketitle

\section{Introduction}

Recently, the Ho\v{r}ava-Lifshitz (HL) gravity has been proposed as an
ultraviolet (UV) completion of general relativity~\cite{horava},
motivated by the Lifshitz theory in the condensed matter physics~\cite{lifshitz}.
The key of the UV completion is the anisotropic scaling between space and time,
\begin{equation}
\label{scaling}
t \to b^{z}\, t, \qquad x^i \to b\, x^i,
\end{equation}
where the Lifshitz parameter $z$ becomes 1 in the infrared (IR) limit.
In general, the HL theory is not invariant under the full
diffeomorphism group of general relativity but under its subgroup,
called the foliation-preserving diffeomorphism; however, in the IR
limit, the full diffeomorphism is somehow recovered.
A mechanism for recovering the full diffeomorphism or the renormalization group flow is yet unsolved issue, but the HL gravity has been intensively studied in the area of cosmology~\cite{ts,kk,mukohyama,brandenberger,ks,ww,mipark,ls,sj} and black hole physics~\cite{lmp,cco,cy,ysmyung,mann,gh,kk:bh,majhi}.

Considering Arnowitt-Deser-Misner (ADM) decomposition~ 
of the metric with 
$ds^2 = - N^2 c^2 dt^2 + g_{ij} (dx^i + N^i dt) (dx^j
+ N^j dt)$~\cite{adm},
the Einstein-Hilbert action can be rewritten as
\begin{equation}
\label{act:EH}
\begin{aligned}
I_{EH} &= \frac{c^3}{16\pi G_N} \int d^4x \sqrt{-\mathcal{G}} \left[ \mathcal{R} - 2\Lambda \right] \\
  &= \frac{c^2}{16\pi G_N} \int dt d^3x \sqrt{g} N \left[ K_{ij} K^{ij} - K^2 + c^2 \left( R - 2\Lambda \right) \right],
\end{aligned}
\end{equation}
where $K_{ij} \equiv \frac{1}{2N} \left[ \dot{g}_{ij} - \nabla_i N_j - \nabla_j N_i \right]$ is the extrinsic curvature of $t= \text{constant}$ hyper-surface, and the dot denotes the derivative with respect to $t$.
Here, $g_{ij}$, $R$, and $\nabla_i$ are the metric, the intrinsic
curvature, and the covariant derivative in the three-dimensional hyper-surface, respectively.
Note that the action~\eqref{act:EH} can be regarded as one of the $z=1$ HL theory and each terms are invariant under the foliation-preserving diffeomorphism.
Then, the first two terms are referred to as kinetic terms, while the other two are potential terms.

In particular, to study the HL gravity,
higher-order potential terms, for instance, $R^3$, will be taken into account.
But there are almost ten possible terms for $z=3$ case, and
the ``detailed balance'' condition can reduce the ten coefficients to three effective ones~\cite{horava}.
Of course, it also implies that 
$(D+1)$-dimensional renormalization can be reduced to the simpler $D$-dimensional renormalization.
By the way, it has been claimed that matter is not UV stable with 
this condition~\cite{calcagni}, and the HL theory with the detailed
balance condition 
does not have the Minkowski vacuum solution.

On the other hand, there have been interesting studies on the cosmological
evolution in the HL gravity  based on the phase space analysis~\cite{ls}.
Authors showed that the cosmological phase changing from the decelerated expansion
to the accelerated expansion is possible assuming the detailed balance
condition and the constant equation-of-state parameter.
However, the first accelerated expansion corresponding to the
inflationary era has not been discussed.
We believe that there should be such a structure if the HL theory is
indeed the UV completion. By the way, the HL gravity itself may be problematic, for instance,
counting its degrees of freedom~\cite{lp,hkg}, description of the asymptotically flat
spacetime~\cite{ks,lmp}, and so on. However, even in spite of these problems, it
deserves to study various aspects since it may give some insight to understand the
quantum gravity. So, 
we would like to investigate whether the smooth cosmological phase transition from the first
accelerated expansion in the early universe to the second accelerated
expansion over the intermediate decelerated
expansion is possible or not. 
 
In section~\ref{sec:hl:cos}, we recapitulate the Ho\v{r}ava-Lifshitz
cosmology and define some relevant quantities without the ``detailed balance'' condition.
Then, the nonlinear equations of motion will be solved using
nonlinear methods in order to study the cosmological behavior in section~\ref{sec:hl:nl}.
By considering the observational data on density parameters, 
we shall plot the phase portrait in section~\ref{sec:hl:pp}, and 
discuss the behavior in the early stage of our universe and its various destiny.
If a dark scalar energy density is assumed to
be negative, then the desired first accelerated expansion appears.
Moreover, we will show that 
it can be smoothly connected with the
second accelerated expansion. 
Finally, discussions  will be given in section~\ref{sec:dis}.


\section{Cosmological setting for $z=3$}
\label{sec:hl:cos}
We now consider the anisotropic scaling~\eqref{scaling} between time and space,
and then $g_{ij}$ and $N$ are invariant while $N^i \to b^{1-z} N^i$ and $c \to b^{1-z} c$.
Requiring that the Planck constant is also invariant, $\hbar\to\hbar$,
we get $E\to b^{-z}E$ and $M\to b^{z-2}M$, where $E$ and $M$ are
energy and mass, respectively.
Then, the kinetic action in the Ho\v{r}ava-Lifshitz gravity (HL) is naturally given as
\begin{equation}
\label{act:kin}
I_\text{kin} = \frac{2}{\kappa^2} \int dt d^3x \sqrt{g} N \left[ K_{ij} K^{ij} - \lambda K^2 \right],
\end{equation}
where $\kappa^2$ is a coupling related to the Newton constant $G_N$, and $\lambda$ is an additional dimensionless coupling constant.
Note that the original kinetic part of Einstein-Hilbert
action~\eqref{act:EH} can be recovered when $\lambda=1$ and $\kappa^2 = 32 \pi G_N/c^2$.
Moreover, the power-counting renormalizability requires $z\ge3$~\cite{horava}.
From now on, we will choose $z=3$ for simplicity, then $\kappa^2$ becomes
dimensionless because of $\kappa^2 \to b^{3-z} \kappa^2$.

Now, the most general potential for $z=3$ can be written as
\begin{equation}
\label{act:pot:3+1}
\begin{aligned}
I_\text{pot} = -\frac{2}{\kappa^2} \int & dt d^3x \sqrt{g} N \Big[ \alpha + \beta R + \gamma_1 R^2 + \gamma_2 R_{ij} R^{ij} + \xi \varepsilon^{ijk} R_{i\ell} \nabla_j R_k^\ell \\
  & + \sigma_1 R^3 + \sigma_2 R R_{ij} R^{ij} + \sigma_3 R_i^j R_j^k R_k^i + \sigma_4 \nabla_i R \nabla^i R + \sigma_5 \nabla_i R_{jk} \nabla^i R^{jk} \Big].
\end{aligned}
\end{equation}
Note that other possible terms are not independent because of the Bianchi identity and symmetries of Riemann tensors.
The potential terms in Eq.~\eqref{act:EH} 
can be recovered when we set
$\alpha = 2\Lambda c^2$, $\beta=-c^2$, and $\gamma_i=\xi=\sigma_i=0$.
Then, the fundamental constants can be identified with
\begin{equation}
\label{fund:const}
c = \sqrt{-\beta}, \qquad G_N = \frac{\kappa^2c^2}{32\pi}, \qquad \Lambda = -\frac{\alpha}{2\beta}.
\end{equation}

On the other hand, the ten coefficients can be reduced to three,
$\mu$, $\Lambda_W$, $\zeta$, 
using the detailed balance condition~\cite{horava}:
\begin{equation}
\label{coeff:dbc}
\begin{aligned}
& \alpha = -\frac{3\kappa^4\mu^2\Lambda_W^2}{16(3\lambda-1)},\ \beta = \frac{\kappa^4\mu^2\Lambda_W}{16(3\lambda-1)},\ \gamma_1 = \frac{\kappa^4\mu^2(1-4\lambda)}{64(3\lambda-1)},\ \gamma_2 = \frac{\kappa^4\mu^2}{16},\ \xi = -\frac{\kappa^4\mu}{4\zeta^2}, \\
& \sigma_1 = \frac{\kappa^4}{8\zeta^4},\ \sigma_2 = -\frac{5\kappa^4}{8\zeta^4},\ \sigma_3 = \frac{3\kappa^4}{4\zeta^4},\ \sigma_4 = -\frac{3\kappa^4}{32\zeta^4},\ \sigma_5 = \frac{\kappa^4}{4\zeta^4},
\end{aligned}
\end{equation}
for which the three-dimensional topologically massive
gravity action~\cite{djt} is given by
\begin{equation}
W = \mu \int d^3x \sqrt{g} (R-2\Lambda_W) + \frac{1}{\zeta^2} \int \chi(\Gamma),
\end{equation}
where $\chi(\Gamma)$ represents the gravitational Chern-Simons term.
However, we will take the general ten coefficients without 
resort to the detailed balance condition~\eqref{coeff:dbc} so that
the total action becomes
\begin{equation}
\label{totalaction}
I=I_\text{kin} +I_\text{pot}+I_\text{mat}
\end{equation}
where we introduced a normal matter action 
$I_\text{mat} $ which is a perfect fluid source of
 the energy density $\rho$ and the pressure $p$.

Now, we are going to consider the Robertson-Walker (RW) metric,
\begin{equation}
\label{met:RW}
ds^2 = - c^2 dt^2 + a^2(t) \left[ \frac{dr^2}{1-kr^2} + r^2 d\Omega_2^2 \right],
\end{equation}
where $k=0,\pm1$ is the normalized spatial curvature.
After some tedious calculations in Eqs.~\eqref{totalaction}, 
the equations of motion can be obtained as
\begin{align}
3 (3\lambda-1) H^2 =&\ \frac{\kappa^2}{2} \rho + 6 \left[ \frac{\alpha}{6} + \frac{k\beta}{a^2} + \frac{2k^2(3\gamma_1+\gamma_2)}{a^4} + \frac{4k^3(9\sigma_1+3\sigma_2+\sigma_3)}{a^6} \right], \notag \\
  =&\ \frac{\kappa^2}{2} \left[ \rho + \rho_\text{vac} + \rho_k + \rho_\text{dr} + \rho_\text{ds} \right], \label{eq:energy} \\
(3\lambda-1) \left( \dot{H} + \frac32 H^2 \right) =&\ -\frac{\kappa^2}{4} p - 3 \left[ -\frac{\alpha}{6} - \frac{k\beta}{3a^2} + \frac{2k^2(3\gamma_1+\gamma_2)}{3a^4} + \frac{4k^3(9\sigma_1+3\sigma_2+\sigma_3)}{a^6} \right], \notag \\
  =&\ -\frac{\kappa^2}{4} \left[ p + p_\text{vac} + p_k + p_\text{dr} + p_\text{ds} \right], \label{eq:pressure}
\end{align}
where $H=\dot{a}/a $ is the Hubble parameter. 
Apart from the normal matter contribution, 
the \emph{additional} energy-momentum contributions from the potential
terms~\eqref{act:pot:3+1} are 
explicitly written as
\begin{equation}
\label{add:en}
\begin{aligned}
  &p_\text{vac} = -\rho_\text{vac} = -\frac{3c^2}{8\pi G_N} \frac{\alpha}{6},
    & \quad & p_k = -\frac13\rho_k = -\frac{c^2}{8\pi G_N} \frac{k\beta}{a^2}, \\
  &p_\text{dr} = \frac13\rho_\text{dr} = \frac{c^2}{8\pi G_N} \frac{2k^2(3\gamma_1+\gamma_2)}{a^4}, 
    & \quad & p_\text{ds} = \rho_\text{ds} = \frac{3c^2}{8\pi G_N} \frac{4k^3(9\sigma_1+3\sigma_2+\sigma_3)}{a^6}.
\end{aligned}
\end{equation}
Similarly to the general relativity, $\rho_\text{vac}$ 
and $\rho_k$ 
come from the vacuum energy and the spatial curvature
contributions, respectively. In particular, in the HL theory, 
$\rho_\text{dr}$ 
is called the \emph{dark radiation}.
Moreover, $\rho_\text{ds}$ 
is  a dark scalar, since it is 
characterized by $\rho_\text{ds} = p_\text{ds}  \sim a^{-6}$.
It is interesting to note that for the spatial flat geometry,
the equations of motion~\eqref{eq:energy} and \eqref{eq:pressure} are
reduced to those of general relativity up to a factor
$(3\lambda-1)/2$; 
in other words, the given HL cosmology is prominent for the 
nonvanishing spatial curvature.
Both $\rho_\text{dr}$ 
and $\rho_\text{ds}$ 
depend on the spatial curvature $k$;
in this sense, these might be regarded as corrections to the curvature contribution $\rho_k$.
The total \emph{actual} energy can be defined by
$\rho_\text{tot}=\rho + \rho_\text{vac}  + \rho_\text{dr} + \rho_\text{ds}$.

From Eqs.~\eqref{eq:energy} and \eqref{eq:pressure}, we can get the
acceleration of the scale factor,
\begin{equation}
\label{eq:acc}
(3\lambda-1) \frac{\ddot{a}}{a} = -\frac{\kappa^2}{12} \left[ \rho_\text{tot} + 3 p_\text{tot} \right],
\end{equation}
where we used the relation $\rho_k+3p_k=0$.
Next, differentiating Eq.~\eqref{eq:energy} with respect to $t$ and plugging it into Eq.~\eqref{eq:pressure}, we can naturally obtain the fluid equation $\dot{\rho}_c + 3 H \left( \rho_c + p_c \right) = 0$,
where $\rho_c$ is defined by $\rho_c \equiv 6(3\lambda-1)H^2/\kappa^2
= \rho_\text{tot} + \rho_k$. Without loss of generality, the
conservation equation becomes
\begin{equation}
\label{eq:fluid}
\dot{\rho}_i + 3 H \left( \rho_i + p_i \right) = 0,
\end{equation}
which is valid for each source component, i.e.,
$\{\rho_i\} = \{\rho, \rho_\text{vac}, \rho_k, \rho_\text{dr}, \rho_\text{ds}\}$.


\section{Linear analyses}
\label{sec:hl:nl}
Ultimately, we want to show the cosmological phase transition from the first
accelerated expansion to the second accelerated expansion
in terms of the intermediate decelerated expansion, which looks like the overall behavior
of our universe. Before we get down to this problem, we are going to exhibit
the essence of the linear analysis.
First, rewriting Eq.~\eqref{eq:acc}, we have
\begin{equation}
\label{eq:acc:eos}
(3\lambda-1) \frac{\ddot{a}}{a} = -\frac{\kappa^2}{12} [1+3\omega(a)] \rho_\text{tot},
\end{equation}
where $\omega(a) =  p_\text{tot}/\rho_\text{tot}$ is the
equation-of-state parameter for the total energy. 
To describe phase portraits, we should make change of variables so that
we introduce $v=\dot{a}$, and
$h=\kappa^2\rho_\text{tot}/6(3\lambda-1)$ 
with $\lambda>1/3$. Then, we can get three first-order differential equations,
\begin{equation}
\label{eq:dyn}
\begin{aligned}
& \dot{a} = v, \\
& \dot{v} = -\frac12 \left[ 1 + 3 \omega(a) \right] a h, \\
& \dot{h} = - 3 \left[ 1 + \omega(a) \right] \frac{v h}{a},
\end{aligned}
\end{equation}
where we used Eqs.~\eqref{eq:fluid} and \eqref{eq:acc:eos}.
Note that integrating Eq.~\eqref{eq:dyn}, we can find a conservation relation,
\begin{equation}
\label{rel:conserv}
a^2h - v^2 = - \frac{2k\beta}{3\lambda-1},
\end{equation}
where the integration constant in the right hand side of \eqref{rel:conserv}
was consistently fixed by comparing Eq.~\eqref{eq:energy}.
Now, let us eliminate $h$ in Eq.~\eqref{eq:dyn} using
the conservation relation~\eqref{rel:conserv}, then we  
have just two first-order differential equations,
\begin{align}
& \dot{a} = v \equiv f(a,v), \label{eq:a} \\
& \dot{v} = -\frac{1}{2a} \left[ 1 + 3 \omega(a) \right] \left( v^2 - \frac{2k\beta}{3\lambda-1} \right) \equiv g(a,v), \label{eq:v}
\end{align}
where the equation-of-state parameter can be written as $\omega(a) =
-1-ah'/3h$ with $h'=dh/da$.

To study the behavior of the scale factor in the phase space, we should calculate the
Jacobian matrix, 
\begin{equation}
\label{mat:A}
A = \left[
\begin{array}{cc}
  \partial_a f & \partial_v f \\
  \partial_a g & \partial_v g
\end{array}
\right]_{(a^*,v^*)} = \left[
\begin{array}{cc}
  0 & 1 \\
  3k\beta\omega'(a^*)/(3\lambda-1)a^* & 0
\end{array}
\right],
\end{equation}
on the fixed point $(a^*,v^*)$
which is easily 
obtained from $\dot{a}=\dot{v}=0$ in Eqs.~\eqref{eq:a} and
\eqref{eq:v},
\begin{equation}
\label{fix:pt}
(a^*, v^*) \text{ satisfying } \left\{
  \begin{aligned}
    & v^*=0, && \text{for } k=0, \\
    & 1+3\omega(a^*)=0, v^*=0, && \text{for } k \ne 0.
  \end{aligned} \right.
\end{equation}
Note that for $k=0$, there are non-isolated fixed points; i.e. every
point on the line $v^*=0$ 
can be a fixed point.
Next, considering an eigenvalue 
equation of $A\mathbf{V}=\eta\mathbf{V}$ which gives
two eigenvalues,
the fixed points can be classified as follows~\cite{strogatz}: 
They are \emph{repellers} if both eigenvalues have positive real
part, \emph{attractors} if both eigenvalues have negative real part.
In particular, they become \emph{saddles} if one eigenvalue is positive and
the other is negative. 
On the other hand, if both eigenvalues are pure imaginary, then they are
\emph{centers}.
Moreover, they are called  \emph{non-isolated fixed
  points} if at least one eigenvalue is zero.

Specifically, the Jacobian matrix~\eqref{mat:A} for $k=0$, is trivial
so that we can find non-isolated fixed points as mentioned earlier, due to the fact
that at least one eigenvalue is zero.
As for $k\ne0$, the Jacobian matrix~\eqref{mat:A} 
becomes
$A=\left[\begin{array}{cc}0&1\\3k\beta\omega'(a^*)/(3\lambda-1)a^*&0\end{array}\right]$, 
and the corresponding eigenvalues are
$\eta=\pm\sqrt{3k\beta\omega'(a^*)/(3\lambda-1)a^*}$ 
where $\lambda>1/3$ is already assumed.
Then, the fixed point~\eqref{fix:pt} becomes a saddle node for $k\beta\omega'(a^*)>0$ or a center for $k\beta\omega'(a^*)<0$. 

\section{Cosmological phase transition}
\label{sec:hl:pp}
We are now in a position to specify the matter source which consists of 
the conventional cold matter and radiation, $\rho=\rho_m+\rho_r$ 
with $\rho_m\sim a^{-3}$ and $\rho_r\sim a^{-4}$.
Let us define $\rho_\text{rad}=\rho_r+\rho_\text{dr}$, following the conventional
way that $\rho_m$ means the density for the cold dark matter as well as baryons.
Then, the equation-of-state parameter $\omega(a)$ is given by
\begin{equation}
\label{eos}
\omega(a) = \frac{3\Omega_\text{ds}+a^2\Omega_\text{rad}-3a^6\Omega_\text{vac}}{3 \left[ \Omega_\text{ds}+a^2\Omega_\text{rad}+a^3\Omega_{m}+a^6\Omega_\text{vac} \right]},
\end{equation}
where the density parameters are defined as $\Omega_i=\rho_i/\rho_c$ 
evaluated at the present universe scale $a_0$, which can be fixed to
$a_0=1$. 
Note that the equation-of-state parameter becomes
$\omega(a_0) \simeq -\Omega_\text{vac}
\simeq-0.7$ at the present scale, since the observational data 
indicates $\Omega_\text{ds},\Omega_\text{rad}\ll\Omega_\text{vac}\simeq0.7$~\cite{wmap5}.

\begin{figure}[pbt]
  \includegraphics[width=0.45\textwidth]{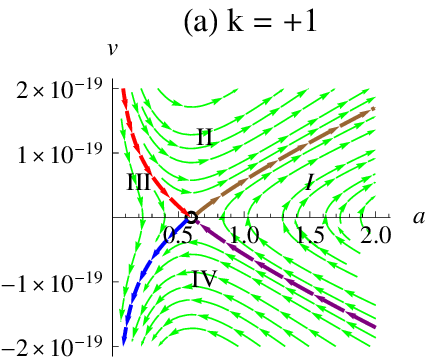}
  \hspace{0.05\textwidth}
  \includegraphics[width=0.45\textwidth]{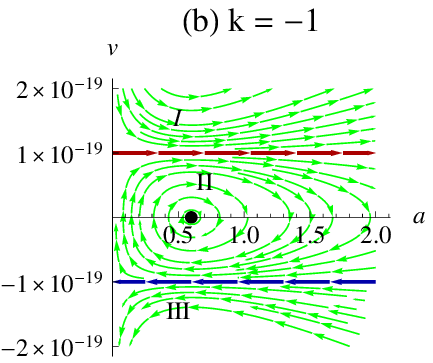}
  \caption{\label{fig:pp}The phase portraits for the configuration of
  our universe are plotted by setting 
 $\Omega_\text{vac}\simeq0.7$, $\Omega_m\simeq0.3$, $\Omega_k\sim10^{-2}k$, $\Omega_\text{rad}\sim10^{-4}$, and $\Omega_\text{ds}\sim-10^{-12}$ for (a) $k=+1$ and (b) $k=-1$.
  The dots on the $a$-axis represent fixed points $(a^*,0)$ such that $1+\omega(a^*)=0$. 
  Actually, these figures looks the same with those of the general relativity. 
  The crucial differences will be explicitly seen in the small scale region in Fig.~\ref{fig:pp:inf}.
}
\end{figure}

In order to see how the desired evolution of the universe appears, we 
plot the phase portraits in Fig.~\ref{fig:pp} by 
setting $\Omega_\text{vac}\simeq0.7$, $\Omega_m\simeq0.3$, $\Omega_k\sim10^{-2}k$, $\Omega_\text{rad}\sim10^{-4}$, and $\Omega_\text{ds}\sim-10^{-12}$.
Explicitly, for $k=+1$ in Fig~\ref{fig:pp}(a), there is a saddle node, and the corresponding stable and
unstable manifolds are shown 
by thick arrows.
It can be also shown that the universe monotonically expanding(II) and
shrinking(IV) phases, and 
the scale factor has a maximum(III)   
if the initial expansion rate is small compared to that of the phase (II).
On the other hand, in the region I in Fig.~\ref{fig:pp}(a), 
the universe can not have a scale less
than that of the fixed point
, i.e. $a>a^*\simeq0.6$, so that 
the shrinking universe starts to expand before the scale factor reaches $a^*$.
For $k=-1$ in Fig.~\ref{fig:pp}(b), there is a center which is a kind of fixed points.
The two straight trajectories separate closed orbits near the
center from the curved trajectories, where the former case describes oscillating
universes(II)
while the latter one describes monotonically expanding(I) or shrinking(III) universes.
Note that the oscillating universe is possible only when
$\rho_\text{tot}<0$, 
and the straight line trajectories are in fact trivial solutions
satisfying $\rho_\text{tot}=0$. 
As a result, the universe can experience late-time accelerated expansion 
after decelerated expansion in both cases (a) and (b), 
because current observational data indicates that the trajectory of
our universe is in the region II in Fig.~\ref{fig:pp}(a) or I in Fig.~\ref{fig:pp}(b).

\begin{figure}[pbt]
  \includegraphics[width=0.45\textwidth]{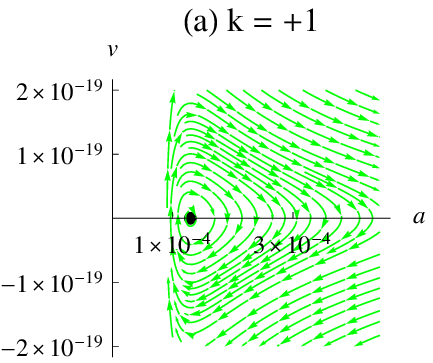}
  \hspace{0.05\textwidth}
 \includegraphics[width=0.45\textwidth]{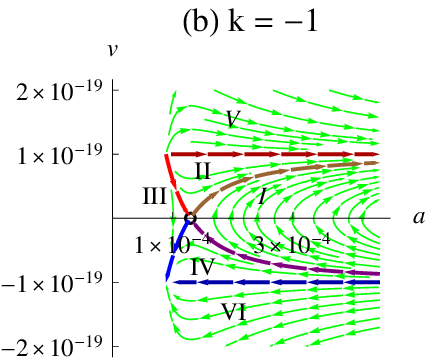}
  \caption{\label{fig:pp:inf}The phase portraits for the configuration of our universe, $\Omega_\text{vac}\simeq0.7$, $\Omega_m\simeq0.3$, $\Omega_k\sim10^{-2}k$, $\Omega_\text{rad}\sim10^{-4}$, and $\Omega_\text{ds}\sim-10^{-12}$, seen near the inflation era for (a) $k=+1$ and (b) $k=-1$. The dots on the $a$-axis represent fixed points $(a^*,0)$ such that $1+\omega(a^*)=0$.}
\end{figure}

Actually, the phase portraits in Fig.~\ref{fig:pp} are very close to
the Einstein's theory in the large scale and
the large expansion rate, so we have to draw small scale
behaviors to expose some differences from the general relativity.
The standard lore tells us that there are no phase transitions in a small scale 
in the Einstein's relativity, unless we consider an additional source
such as the inflaton.
However, it is possible to see a transition with the help of the dark scalar in the HL theory, which is shown in Fig.~\ref{fig:pp:inf}.
The dark scalar looks like a quantum correction, 
in the sense that it comes from higher curvature correction and 
it plays a significant role at the small scale, $a\ll1$. 
So, if we assume $\Omega_\text{ds}<0$, then we can have another fixed point near $a^*\sim10^{-4}$,
which gives the possibility for the first accelerated expansion.
For the case of $k=+1$ in Fig.~\ref{fig:pp:inf}, there is a center near the early accelerated era.
One can now see that the expanding and 
shrinking universe in the region III in Fig.~\ref{fig:pp}(a) is actually
oscillating, which means a universe with the same energy distribution 
as ours should oscillate if its initial expansion rate is not
so large enough.  
As for the case of $k=-1$ in Fig.~\ref{fig:pp:inf}, there is a saddle node with
stable and unstable manifolds, 
and the two straight trajectories whose total energy is trivial,
$\rho_\text{tot}=0$ which separates solutions satisfying
$\rho_\text{tot}>0$ from $\rho_\text{tot}<0$ solutions. 

\begin{figure}[pbt]
  \includegraphics[width=0.45\textwidth]{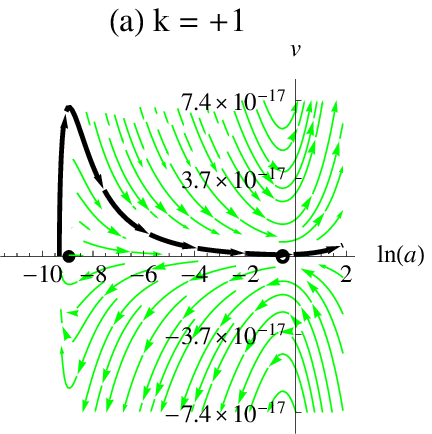}
  \hspace{0.05\textwidth}
  \includegraphics[width=0.45\textwidth]{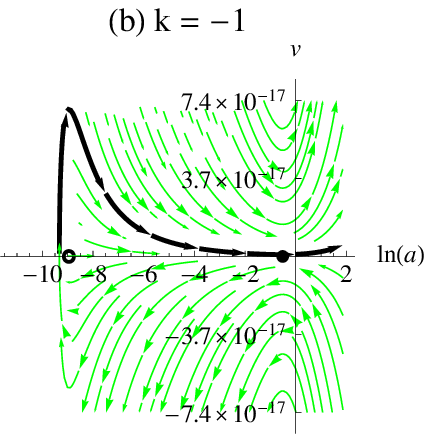}
  \caption{\label{fig:pp:tot}The phase portraits for the configuration of our universe, $\Omega_\text{vac}\simeq0.7$, $\Omega_m\simeq0.3$, $\Omega_k\sim10^{-2}k$, $\Omega_\text{rad}\sim10^{-4}$, and $\Omega_\text{ds}\sim-10^{-12}$, for (a) $k=+1$ and (b) $k=-1$. The thick black (curved) arrows describe the trajectory of our universe, and the dots on the $a$-axis represent fixed points $(a^*,0)$ such that $1+\omega(a^*)=0$.}
\end{figure}

To incorporate Fig.~\ref{fig:pp} with Fig.~\ref{fig:pp:inf}, they are plotted
in terms of the logarithmic scale in Fig.~\ref{fig:pp:tot},
and then it can be shown that the universe starts from the first accelerated phase 
followed by a decelerated expansion phase, and ends up
with the second accelerated expansion.
In addition to this dynamical evolution, as seen from the two fixed
points in Fig.~\ref{fig:pp:tot}, it is also possible to obtain
the neutrally stable universe described by the center at the early stage of universe in
Fig.~\ref{fig:pp:tot}(a) and at the late time stage of universe in Fig.~\ref{fig:pp:tot}(b).
Here, `neutral' means that a small perturbation does not decay into zero
but remains small as time goes on.

\section{Discussion}
\label{sec:dis}
We have studied the HL gravity coupled to the 
matter without the detailed balance condition 
in order to show the possibility to get the smooth phase transition from 
the first accelerated expansion corresponding to the early stage of the
universe to the second accelerated expansion throughout the intermediate 
decelerated expansion assuming the energy density for the dark scalar
to be negative $\rho_\text{ds}<0$.
In spite of this negative energy contribution, 
the total energy
density  $\rho_\text{tot}$ is positive at any cosmological scale.
Note that there have been researches which provide concrete
justifications for models with negative density, in particular, a brane
universe moving in a curved higher dimensional bulk space~\cite{kk:mirage} and 
a model of dark energy stemming from a fermionic condensate~\cite{abc}.
Of course, we can confirm that there is no such a phase transition
unless the dark scalar density is assumed to be negative. 
For instance, if we take the detailed balance condition, then it
simply reduces the additional energy contribution 
to $\rho_\text{vac} = \Lambda c^4 / 8\pi G_N$, $\rho_k = -3c^4k / 8\pi G_N a^2$, $\rho_\text{dr} = 27c^4k^2 / 32\pi G_N \Lambda a^4$, $\rho_\text{ds} = 0$
as seen from fundamental constants~\eqref{fund:const} and
redefinition of coefficients~\eqref{coeff:dbc}. In this case, the
first accelerated expansion does not appear, that is the reason why
we did not take the detailed balance condition.
In addition, it has been shown that our universe 
may be oscillatory even with the same density parameters of the
current observation, depending on the expansion rate. 
Actually, the expansion rate is related to the total energy density 
by Eq.~\eqref{rel:conserv}, so that small expansion rate $v$
corresponds to small $h\sim\rho_\text{tot}$ for a given scale factor.

On the other hand, it has been well known that if there were $\sim60$ e-foldings of inflationary
expansion, then the universe was driven to be nearly flat, removing
any need for fine-tuned initial conditions. In addition, the inflationary
expansion would have driven the density of magnetic monopoles to
be negligible today, explaining their apparent absence. Also, our
entire observable portion of the universe would have inflated from an
initially small causally connected region, thereby ensuring a high degree
of isotropy today. Although our analysis shows that the first accelerated expansion can be
obtained by assuming $\rho_\text{ds}<0$, unfortunately, it should be pointed out
that it does not last for 60 e-foldings.
We hope this problem will be discussed elsewhere.

It is interesting to note that 
there exists a minimum scale $a_\text{min}$ 
obtained from $\rho_c\ge0$ and $\rho_\text{tot}\ge0$ with the relation $\rho_c=\rho_\text{tot}+\rho_k$.
Note that $\rho_\text{tot}(a_\text{min})=-\rho_k(a_\text{min})>0$ and
$H^2(a_\text{min})\sim\rho_c(a_\text{min})=0$ for $k=+1$, so the
universe starts with a certain amount of energy and zero expansion
rate, while $\rho_\text{tot}(a_\text{min})=0$ and
$H^2(a_\text{min})\sim \rho_c(a_\text{min})=\rho_k(a_\text{min})>0$
for $k=-1$, so the universe starts with zero energy density and a
certain expansion rate. For any case, 
as far as we consider the positive total energy density
and the critical energy density, there should be no initial
singularity problem 
because of the existence of the minimum scale.

The final comment is in order. 
The energy density and the pressure of the dark scalar which play a crucial role
in this work depend on the three terms in the potential~\eqref{act:pot:3+1}, 
$\sigma_1 R^3$, $\sigma_2 R R_{ij}
R^{ij}$, $\sigma_3 R_i^j R_j^k R_k^i$.   
One may think that three terms give rise to kinetic
contributions so that they provide the leading spatial dependence in
the gravition propagator in the UV region. What it means is 
that graviton modes may be unstable.
However, this is not the case 
since they do not modify the gravition propagator.
In general, the lowest metric order in curvature tensors and curvature
scalars is linear, which means that the three terms are not
quadratic in metric. Of course, such terms which are quadratic in curvature, 
$\sigma_4 \nabla_i R \nabla^i R$, $\sigma_5 \nabla_i R_{jk} \nabla^i
R^{jk}$ in Eq.~\eqref{act:pot:3+1}, will not only add interaction but also modify the
propagator; however, these are irrelvant to our energy density formulae~\eqref{add:en}.


\begin{acknowledgments}
This work was supported by the National Research Foundation of Korea(NRF) grant funded by the Korea government(MEST) through the Center for Quantum Spacetime(CQUeST) of Sogang University with grant number 2005-0049409. W.~Kim was also supported by the Special Research Grant of Sogang University, 200911044.
\end{acknowledgments}


\end{document}